\documentclass[10pt,twocolumn]{article}

\usepackage[]{amsmath}
\usepackage[]{mathtools}
\usepackage[]{wasysym}
\usepackage{lineno}
\usepackage[a4paper, portrait, margin=1in]{geometry}
\usepackage[backend=bibtex,style=numeric-comp,sorting=none,maxbibnames=99]{biblatex} 
\begin{document}

\onecolumn

\begin{center}
 
\vspace*{0.2\baselineskip \LARGE PFG NMR time-dependent diffusion coefficient analysis of confined emulsion: post drainage phase conformation.  \linebreak}

\vspace*{0.5\baselineskip \large B. Chencarek$^{a,*}$, M. Nascimento$^{a,b}$, A. M. Souza$^{a}$, R. S. Sarthour$^{a}$, B. Coutinho$^{b}$, M. D. Correia$^{b}$, I. S. Oliveira$^{a}$}

\vspace*{1.5\baselineskip \normalsize \textit{$^{a}$Centro Brasileiro de Pesquisas F\'isicas, Rua Dr. Xavier Sigaud 150 Ed. Cesar Lattes,
	Urca, Rio de Janeiro, RJ. CEP 22290-180, Brasil}}

\vspace*{0.5\baselineskip \normalsize \textit{$^{b}$Centro de Pesquisas e Desenvolvimento Leopoldo Am\'erico Miguez de Mello - CENPES /PETROBRAS, Av. Hor\'acio Macedo, 950, Cidade Universit\'aria, Rio de Janeiro, RJ. CEP 21941-915, Brasil}}

\vspace*{0.5\baselineskip \normalsize \textit{$^{*}$Corresponding author: bruno.chencarek@cbpf.br / Phone: +552121417112}}

\vspace*{1\baselineskip \rule{42em}{0.2ex}}

\end{center}

\section*{Abstract}

In this work, we present a characterization of phase configuration in water-saturated sintered glass bead samples after oil injection, through the analysis of time-dependent diffusion coefficients obtained from sets of one-dimensional pulsed field gradient nuclear magnetic resonance (PFG NMR) measurements, pre and post drainage. Estimates of samples surface-to-volume ratio and permeability from pre drainage PFG measurements in a water-saturated sample were compared with analytical and reported values, respectively, and a fair agreement was found in both cases. Short-time analysis of diffusion coefficients extracted from PFG measurements was used to quantify the increase in surface-to-volume ratio probed by the wetting phase after drainage. Analysis of water and oil diffusion coefficients from post drainage PFG experiments were carried out using a bi-Gaussian model, and two distinct scenarios were considered to describe fluids conformation within pores. For the case where non-wetting phase was considered to exhibit a poorly connected geometry, an analysis assuming the formation of oi-in-water droplets within pores was performed, and a Gaussian distribution of droplets radii was determined.\par

\vspace*{1\baselineskip \textit{Keywords}: PFG NMR; emulsion; drainage; injection; porous media; petrophysics}

\begin{center}
\vspace*{0.5\baselineskip \rule{42em}{0.2ex}}
\end{center}

\pagebreak

\twocolumn

\section{Introduction}

Characterization of emulsions (immiscibe fluid mixtures) is an essential part of research in different industry segments such as foods, chemicals, pharmaceutics and also oil exploration \cite{GUZEY2006EmulsionFood,Khan2011PharmaReview,LAWRENCE2000Pharma,SpeightEmulsionReview,PERAZZO2018Emulsion}. For the latter, due to the very own nature of oil formation and migration processes, reservoir rocks are commonly found saturated with water-oil mixtures, and some oil recovery strategies also utilize emulsions injection in underground reservoirs as a mechanism to increase oil extraction \cite{McAuliffe1979Migration,PERAZZO2018Emulsion}. Among several different tools used to measure physical and chemical properties from fluids in saturated rocks, NMR relaxometry techniques stand out as being non-invasive, and have been long used in porous media petrophysics \cite{SerraWellLoggingBook, NMRLoggingBook}.\par 

Although application of one-dimensional relaxometry techniques and their theoretical framework on rocks saturated with a singe fluid allows a relatively straightforward data interpretation, analysis of relaxation profiles becomes problematic when porous space contains more than one phase. Water and oil subjected to different confinement conditions may exhibit similar relaxation rates, depending on pore features like geometry and surface physico-chemical properties \cite{dunn2002nuclear,Muncaci}. Moreover, as a result of magnetization transfer between present phases, exchange effects must now be accounted for into the Bloch equation formalism  \cite{McConnellExchange} and the behavior of solutions can be become quite distinct from what it is generally expected from the single phase case. \par

A significant improvement in fluid identification can be achieved by combining molecular diffusion measurements, obtained from pulsed-field gradient (PFG) NMR, and relaxometry protocols into two-dimensional techniques such as diffusion-relaxation (D-$T_{2}$) \cite{ZHANG2014DT2} and diffusion-diffusion (D-D) \cite{CallaghanDD} correlation maps. Despite the considerable increase in available information, these experiments can be very time-consuming and the extraction of correlation maps from raw data often relies on two-dimensional inverse Laplace Transform algorithms \cite{Venkataramanan2002Laplace2D}.\par 

Restricted diffusion coefficients obtained from PFG NMR experiments in one dimension also provide valuable information on bulk emulsion \cite{PACKER1972,HOLLINGSWORTHEmulsionPFG} and, for the case of molecules undergoing diffusion in the pore space, parameters pertaining to the confining geometry, such as porous media surface-to-volume ratio, become accessible under specific diffusion regimes \cite{diffusioNMRconfinedBook,MitraShortTime,HURLIMANN1994}. \par

In this work we present the analysis of time-dependent diffusion coefficients, obtained from sets of one-dimensional PFG NMR measurements, to characterize the effects of drainage in phase conformation in water-saturated sintered glass beads, in order to identify and evaluate individual characteristics of water and oil phases, and mixture conformation features after drainage.\par 

The present work is organized as follows. First a brief overview on distinct scenarios for fluids conformation in drainage experiments is introduced in Section 2. In Section 3 a theoretical background for the analysis of molecular diffusion in confined systems is presented. Samples preparation and experimental protocols are described in Section 4. In Section 5 we present the results obtained from PFG NMR measurements, and an analysis on fluids dynamics. This section is divided into three parts, containing bulk fluids characterization, and a PFG NMR analysis on pre and post drainage measurements, respectively. Conclusions are presented in Section 6.

\section{Fluid conformation after drainage}

When oil is forced into a porous rock previously saturated with water, an emulsion can be formed inside the pore space. Conformation of fluids within pores depends on several factors such as physico-chemical properties of both phases, injection pressure, pores connectivity, and also on the relation between wetting and non-wetting phases \cite{HirasakiWettability}. Wettability is the property that intermediates the contact between a liquid and a solid surface, determined by the interplay among attractive (adhesive) and repulsive (cohesive) forces. Although surface wettability can be altered by physico-chemical processes in the long-time presence of two fluids, reservoir rocks are commonly water-wet \cite{PERAZZO2018Emulsion}.\par 

\begin{figure}[!ht]
	\centering
	\includegraphics[width=220pt]{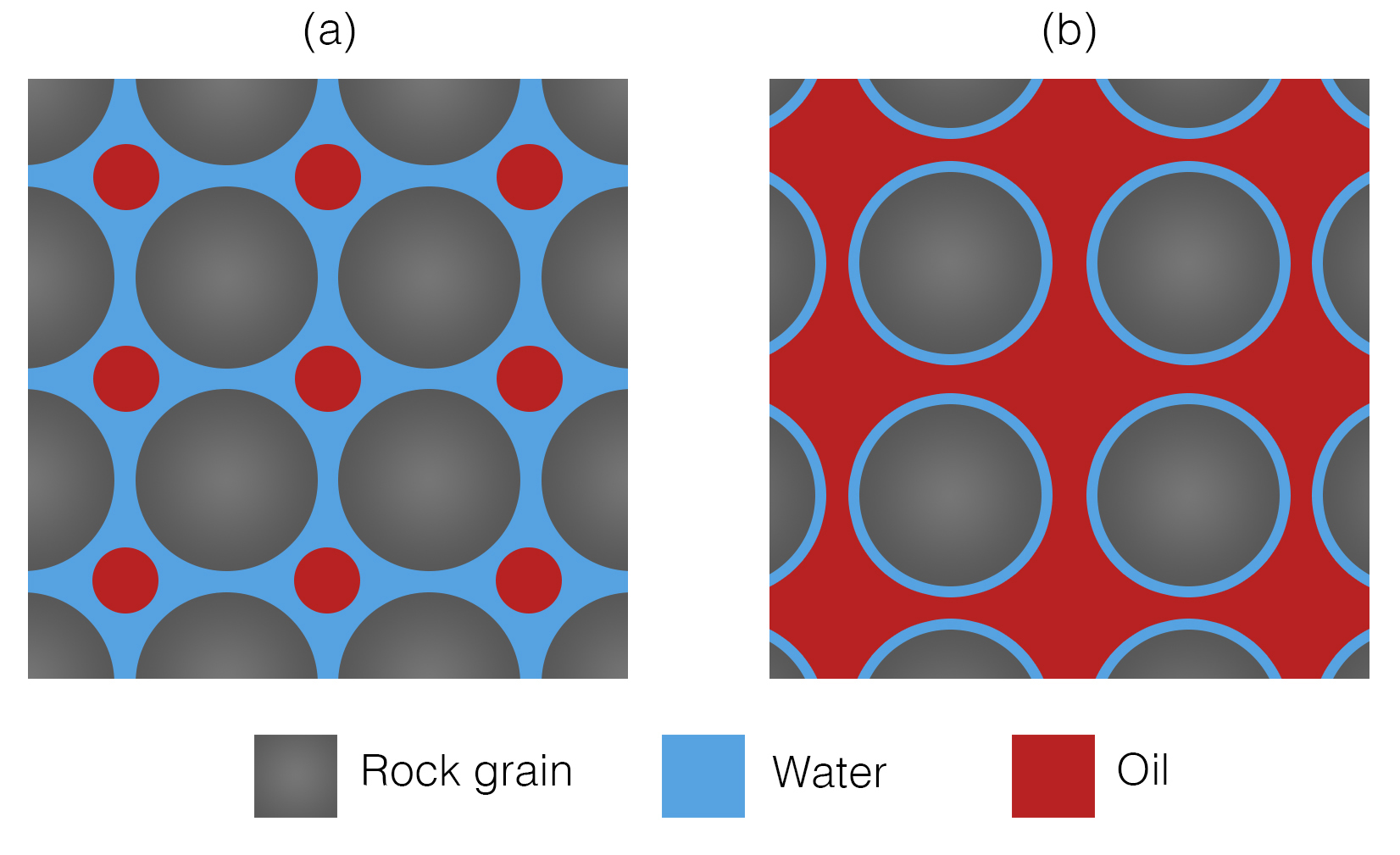}
	\caption{Illustration of possible conformation scenarios for oil injection in water saturated porous media. (a) Oil is located only in the innermost region of pores, with pore throats still filled with water. In this scenario the non-wetting phase exhibits limited or no connectivity. (b) Non-wetting phase fills both inner regions of pores and throats, in a highly-connected configuration.}
	\label{fig_injection}
\end{figure}

Oil injection in a water-saturated porous sample, on what regards conformation and connectivity of wetting and non-wetting phases, may result into two distinct scenarios. In the first one (Fig. \ref{fig_injection} - (a)) oil is distributed into several droplets, located in the innermost region of pores. In this case pore throats remain filled with water, and the non-wetting phase exhibits limited or no connectivity. In a second scenario oil fills both inner regions of pores and throats, in a highly-connected configuration (Fig. \ref{fig_injection} - (b)).

Conformation of water and oil phases after drainage and porous geometry features will determine how molecules of each fluid diffuse through the pore space. Molecular movement is now restricted and self-diffusion coefficients measured in PFG NMR experiments reflect these constraints. In the next section a theoretical background for molecular diffusion and its application in confined systems will be presented. 

\section{Molecular diffusion in confined systems}

%
%
%
%

\begin{figure}[!h]
 	\centering
 		\includegraphics[width=220pt]{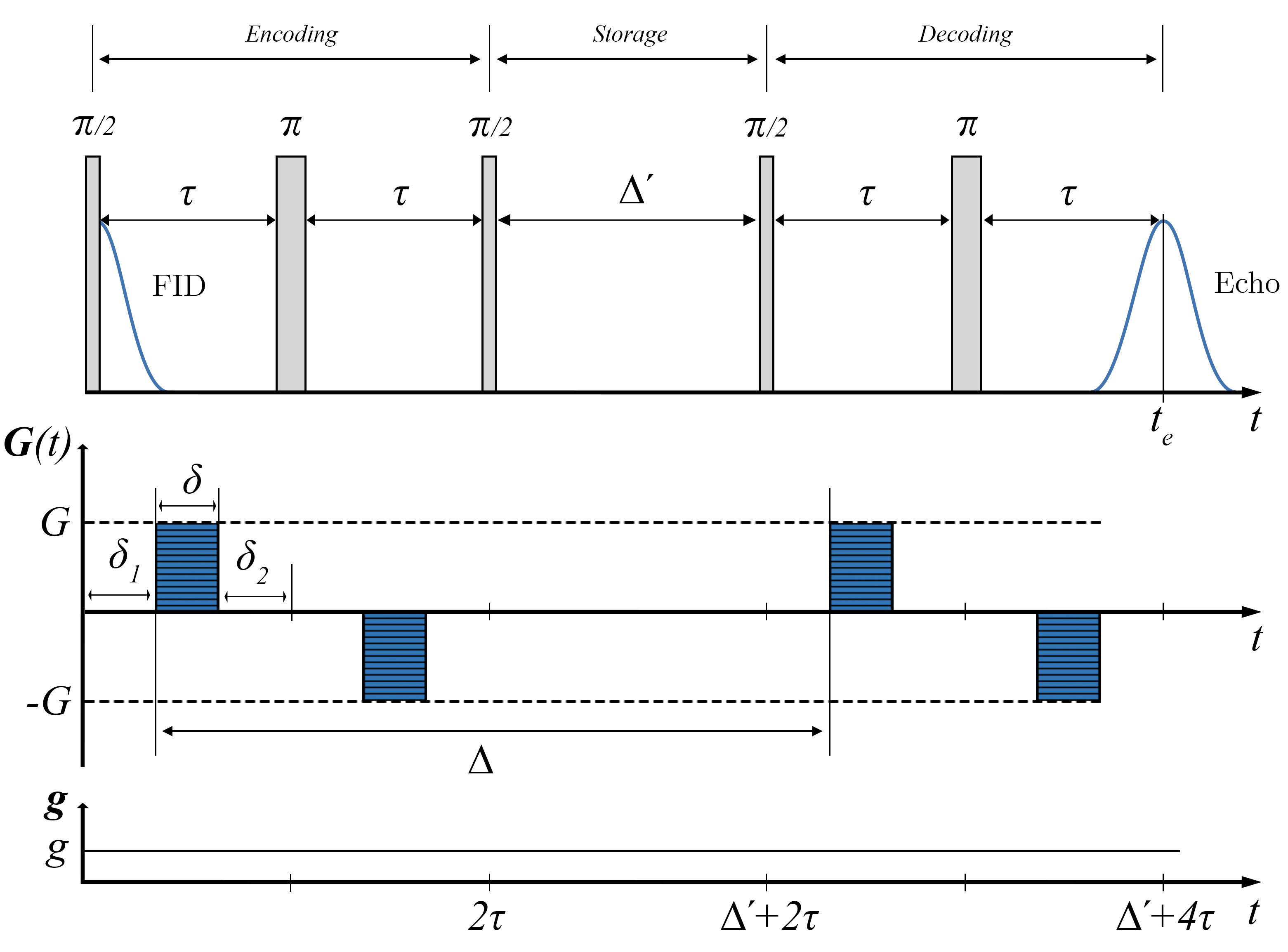}
 		\caption{13-interval PFG NMR pulse sequence proposed by Cotts et al. in the presence of pulsed magnetic field gradients ($G$) and a constant background gradient ($g$). In the radio-frequency pulses (upper scheme) $t_{e}$ denotes the echo formation time.}
 		\label{fig_seq_Cotts}
 \end{figure}

The NMR signal attenuation due to molecules self-diffusion $\Psi(t)$ can be calculated solving the double integral \cite{GalvosasReviewDiffusion}:
\begin{equation}
\label{dif_attenuation}
	\Psi(t) = exp\Big\{-D\gamma^2\int_{0}^{t}dt'\Big[\int_{0}^{t'}dt''G^{*}(t'')\Big]^{2}\Big\}, 
\end{equation} 

\noindent wherein $D$ is the molecule self-diffusion coefficient, $\gamma$ is the gyromagnetic ratio and $G^{*}(t)$ represents the effective magnetic field gradient. In this work all diffusion measurements were carried out using a 13-interval PFG NMR sequence (Fig. \ref{fig_seq_Cotts}) proposed by Cotts et al. \cite{COTTS1989_13_interval}, in the presence of pulsed ($G$) and background ($g$) magnetic field gradients. Integration in Equation (\ref{dif_attenuation}) taking the effective gradient as $G^{*}(t) = G(t) + g(t)$ leads to \cite{GalvosasReviewDiffusion}:

\begin{equation}
\label{dif_attenuation_sol}
\Psi(t_{e}) = exp\{-D\gamma^2[A_{p}(t_{e}) + A_{b}(t_{e}) + A_{c}(t_{e})]\},
\end{equation}

\noindent wherein, $A_{p}$, $A_{c}$ and $A_{b}$ denote pulsed, cross and background gradient terms, respectively, and for the pulse sequence represented in Figure \ref{fig_seq_Cotts} we have:
\begin{equation}
\label{dif_attenuation_sol_pulsed}
A_{p}(t_{e}) = (2\delta)^2\Big[\Delta'+\frac{3}{2}\tau-\frac{1}{6}\delta\Big]G^2,
\end{equation}
\begin{equation}
\label{dif_attenuation_sol_cross}
A_{c}(t_{e})=2\delta\tau(\delta_{1}-\delta_{2})Gg
\end{equation}
\noindent and
\begin{equation}
\label{dif_attenuation_sol_back}
A_{b}(t_{e}) = \frac{4}{3}\tau^3g^2.
\end{equation}
The cross gradient term $A_{c}$ vanishes when $\delta_{1} = \delta_{2}$. If pulsed gradients $G$ are much larger than constant background gradient $g$, and the condition $g\tau \ll G\delta$ is fulfilled, the background term $A_{b}$ can also be excluded. Considering this approximation, diffusion attenuation expression is reduced to: \par
\begin{equation}
\label{dif_attenuation_exp}
\Psi(t_{e}) = exp\Big\{-D\gamma^2 (2\delta)^2\Big[\Delta'+\frac{3}{2}\tau-\frac{1}{6}\delta\Big]G^2\Big\}.
\end{equation}

Brownian motion of fluid molecules undergoing unrestricted diffusion (bulk fluid) is described by a normal distribution, and the diffusion propagator is, in this case, a Gaussian function. The signal attenuation measured by PFG NMR can be adjusted by Equation (\ref{dif_attenuation_exp}) and the molecular diffusion coefficient $D$ can be obtained directly proportional to the slope of a $ln[\Psi(t)]$ versus $b$ plot, where parameter $b$ in this case represents:
\begin{equation}
\label{eq_b}
b =  \gamma^{2}(2\delta)^2\Big[\Delta'+\frac{3}{2}\tau-\frac{1}{6}\delta\Big]G^2.
\end{equation}
For the case of confined fluids, molecular movement is now restricted by pore geometry, and this constraint affects the probability density of finding a particle in a given position. As a consequence, the diffusion propagator is no longer a Gaussian function \cite{diffusioNMRconfinedBook}. However, the influence of pore geometry on the diffusive motion depends on the balance between how much time molecules have to propagate and the pore confinement scale. If the diffusion time is short, only molecules that are closer to pore walls will have their movement restricted. For the rest of them, motion goes unrestricted. In this case, if the diffusion length is much smaller than characteristic confinement scale, the propagator can be approximated by a Gaussian function. The validity of this approximation for fluid diffusion in confined systems, known as short-time regime, may be expressed by the condition: 

\begin{equation}
\label{short_time_cond}
\sqrt{D_{0} \Delta} \ll R_{p},
\end{equation}    

\noindent in which $D_{0}$ is the bulk self-diffusion coefficient, $\Delta$ denotes the diffusion (or storage) time and $R_{p}$ represents the characteristic pore radius.\par 

For molecules undergoing restricted motion, the diffusion coefficients obtained by PFG NMR reflect those geometry-imposed restrictions, and their measured values will depend on the diffusion time ($D = D(\Delta)$). Mitra et al. \cite{MitraShortTime} presented an analytical expression for the behavior of time-dependent diffusion coefficients $D(t)$ in the short-time regime as a function of confinement geometrical parameters. For pore walls with smooth boundaries in the presence of surface relaxativity, $D(t)$ can be expressed by: 

%
%
%

\begin{equation}
\begin{aligned}
\frac{D(t)}{D_{0}} = {} & 1-\frac{4(D_{0}t)^{1/2}}{9\sqrt{\pi}}\frac{S}{V} \\
                                  & + (D_{0}t)\frac{S}{V}\Bigg[-\frac{H}{12} + \frac{\rho}{6D_{0}}\Bigg] + \mathcal{O}[(D_{0}t)^{3/2}],
\end{aligned} 
\label{eq_Mitra}
\end{equation}

\noindent wherein ($S/V$) denotes the porous media surface-to-volume ratio, $H$ is the pore mean curvature, $\rho$ represents the surface relaxativity and the last term on the right side stands for the contributions of the order of $(D_{0}t)^{3/2}$. For long diffusion times ($\sqrt{D_{0} \Delta} \gg R_{p}$), molecules will, on average, probe the pore matrix connectivity and the ratio $D(t)/D_{0}$ can be related with porous media diffusive tortuosity\footnote{Different tortuosity definitions can be found in literature related to porous media hydraulic, electrical, geometrical and diffusive connectivity. Although under certain conditions some of these definitions may represent the same property, a distinction between them must be made. An enlightening discussion on different tortuosity definitions and their applications was presented by Ghanbarian et al. \cite{tortuosityReview2013}. Here, the authors have adopted the diffusive tortuosity definition as the parameter related to long-time diffusion behavior. } ($\tau_{d}$) by the relation \cite{Pileio2018SingletTortuosity}:

\begin{equation}
\label{eq_tortuosity}
\lim_{t\to\infty}\frac{D(t)}{D_{0}} =  \frac{1}{\tau_{d}}.
\end{equation} 

\section{Experimental description}

\subsection{Samples and preparation}

\begin{figure}[h!]
	\centering
	\includegraphics[width=220pt]{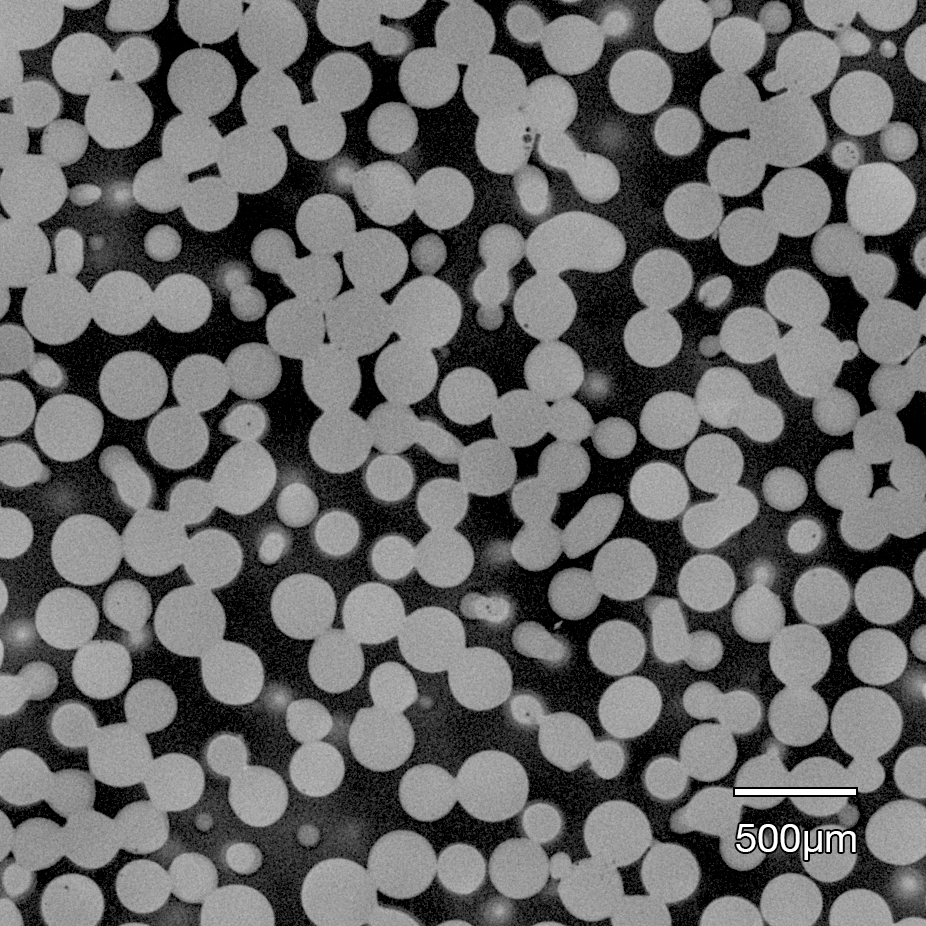}
	\caption{2D micro-tomography image of microspheres after the sintering process employed in the fabrication of the synthetic porous samples. The sintering protocol was designed to preserve most of the porosity from the original package \cite{Chencarek2019}.}
	\label{fig_MicroCT}
\end{figure}

Samples of distilled water ($100\%$) and isoparaffin (ISOFAR 17/21 $100\%$) were used in the experiments. Synthetic porous samples were produced by glass microspheres (diameter $250-300\mu$m) sintering. The fabrication protocol is the same as described in Ref. \cite{Chencarek2019}. The microspheres were placed in a cylindrical ceramic crucible ($Al_{2}O_{3}$ $99.8\%$) and taken to a chamber furnace (Carbolite CWF 1200) for the following heat treatment: from room temperature to 560 C$^\circ$ at 140 C$^\circ/$min, held for 20 minutes; from 560 to 700 C$^\circ$ at 20 C$^\circ$/min, held for 1 hour; from 700 to 490  C$^\circ$ at 14  C$^\circ$/min, 490 to 440  C$^\circ$ at 28 C$^\circ$/min, and then cooled to room temperature. The effect of the thermal treatment, designed to lightly fuse the microspheres while preserving most of the porosity of the original sphere package before sintering, can be seen in the two-dimensional micro-tomography image presented Figure \ref{fig_MicroCT}. 

Water saturation was performed by imbibition using a desiccator connected to a vacuum pump. The dry sample was placed inside a desiccator containing a recipient with distilled water. Only after the establishment of a low-vacuum condition the sample is then dropped into the water container in order to prevent the formation of trapped air bubbles within pores. Samples were left submerged in the water recipient in vacuum condition for approximately 30 minutes to remove residual air content. 

\begin{figure}[h!]
	\centering
	\includegraphics[width=220pt]{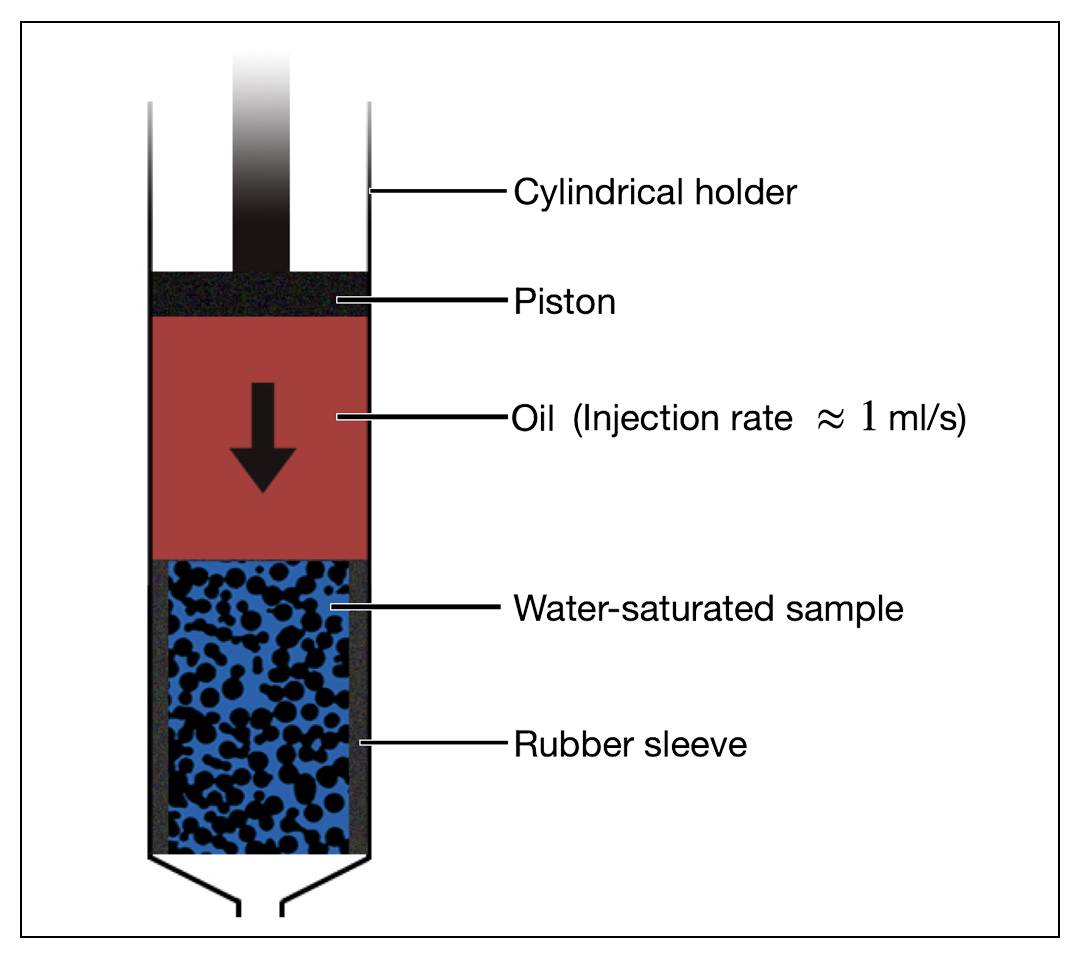}
	\caption{Illustration of the experimental setup employed for the drainage of water-saturated samples. An oil volume equal to the sample's total volume was forced into the water-saturated samples using a manually controlled piston at an average rate of 1 ml/s.}
	\label{fig_drainage}
\end{figure}

For the drainage process the water-saturated samples were clothed in a rubber sleeve and placed into a cylindrical holder, illustrated in Figure \ref{fig_drainage}. An oil volume equal to the sample's volume was manually forced into the sample using a piston at an average injection rate of 1 ml/s. Immediately after the drainage process samples were placed in the glass tubes for the NMR measurements. 

\subsection{Measurements}   

$^{1}$H Spectroscopy, inversion recovery ($T_{1}$) and PFG NMR measurements were performed at room temperature (298.15 K) in bulk samples of both fluids, in an oil-saturated porous sample, and in a water-saturated porous sample before and after oil injection. Measurements were carried out in a $11.7$ T spectrometer (Varian/Agilent) using a 5 mm probe.\par 

A $13$-interval bipolar PFG NMR sequence (Fig. \ref{fig_seq_Cotts}) was used for diffusion measurements. Acquisitions were performed fixing the storage time $\Delta$ and varying the magnetic field gradient strength to monitor the signal attenuation due to diffusion. In this protocol, known as time-independent acquisition, signal loss due to relaxation at echo time formation $R(t_{e})$ is the same for all points in the magnetic field gradient strength array, and in this case, it becomes a normalization constant. This type of acquisition is specially useful for extracting diffusion coefficients of confined fluid mixtures, since the calculation of relaxation attenuation effects in such conditions can be considerably burdensome.\par 

PFG NMR experiments were performed with 18 different diffusion times $\Delta$ varying from 3 to 60 ms. The duration of magnetic field gradient pulses $\delta$ was set as 1 ms, and $\tau$ duration was set as 1.4 ms for all experiments.

\section{Results and discussion}

\subsection{Fluids characterization}

\begin{figure*}[]
	\begin{center}
		\includegraphics[width=\textwidth]{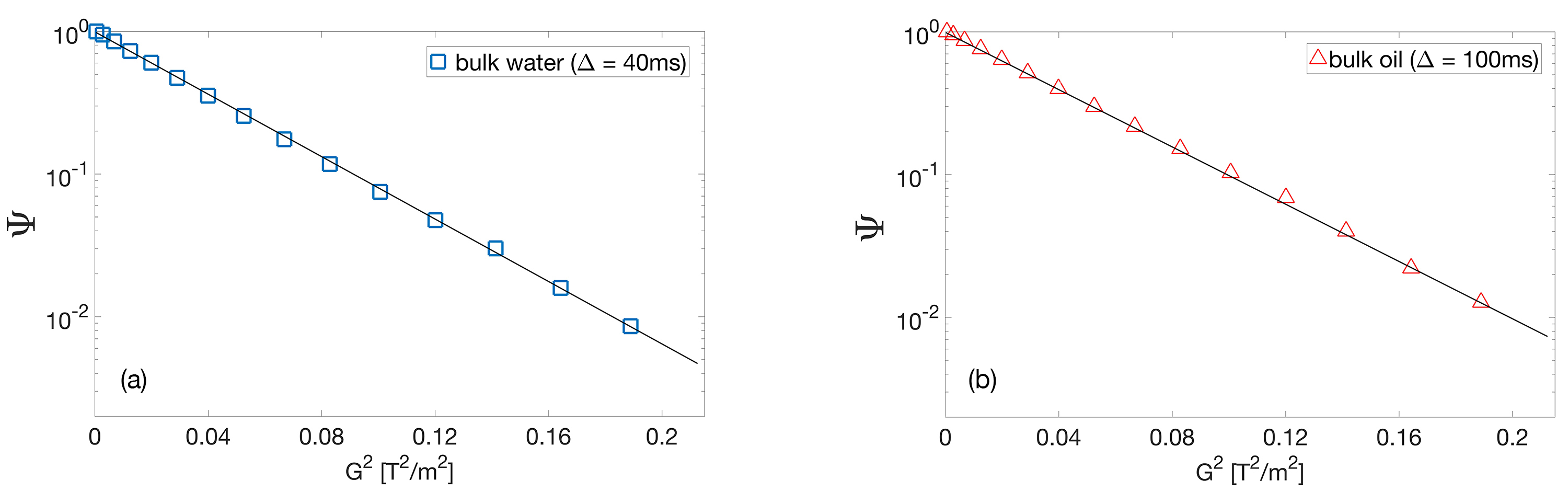}
		\caption{Bulk water (a) and oil (b) normalized spin echo attenuation $\Psi$ as a function of squared gradient $G^{2}$. Diffusion time ($\Delta$) value was 40 ms for water and 100 ms for oil measurements. Black lines denote fittings using Equation (\ref{dif_attenuation_exp}). Self-diffusion coefficients obtained were $D^{w}_{0} = 2.29 \times 10^{-9}$ m$^{2}$/s for water and $D^{o}_{0} = 0.79 \times 10^{-9}$ m$^{2}$/s for oil.}
		\label{fig_dif_water_oil_bulk}
	\end{center}
\end{figure*}

Individual phase self-diffusion coefficients were measured in bulk samples of each fluid. Figure \ref{fig_dif_water_oil_bulk} shows signal attenuation $\Psi$ as a function of squared gradient $G^{2}$, and linear fittings performed with Curve Fitting toolbox (MATLAB version R2019b), using the expression in Equation (\ref{dif_attenuation_exp}). Obtained values were $D^{w}_{0} = 2.29 \: 10^{-9}$ m$^{2}$/s for water (left plot) and $D^{o}_{0} = 0.79 \: 10^{-9}$ m$^{2}$/s for oil (right plot). The fitting error was no greater than 1 $\%$ for both samples and the water diffusion coefficient value is in fair agreement with previously reported values \cite{MillsWaterDiffCoeff}. No reference diffusion coefficient value was found in literature for this particular oil sample at 298.15 K.

\subsection{Pre drainage PFG analysis}

\begin{figure*}[!ht]
	\begin{center}
		\includegraphics[width=\textwidth]{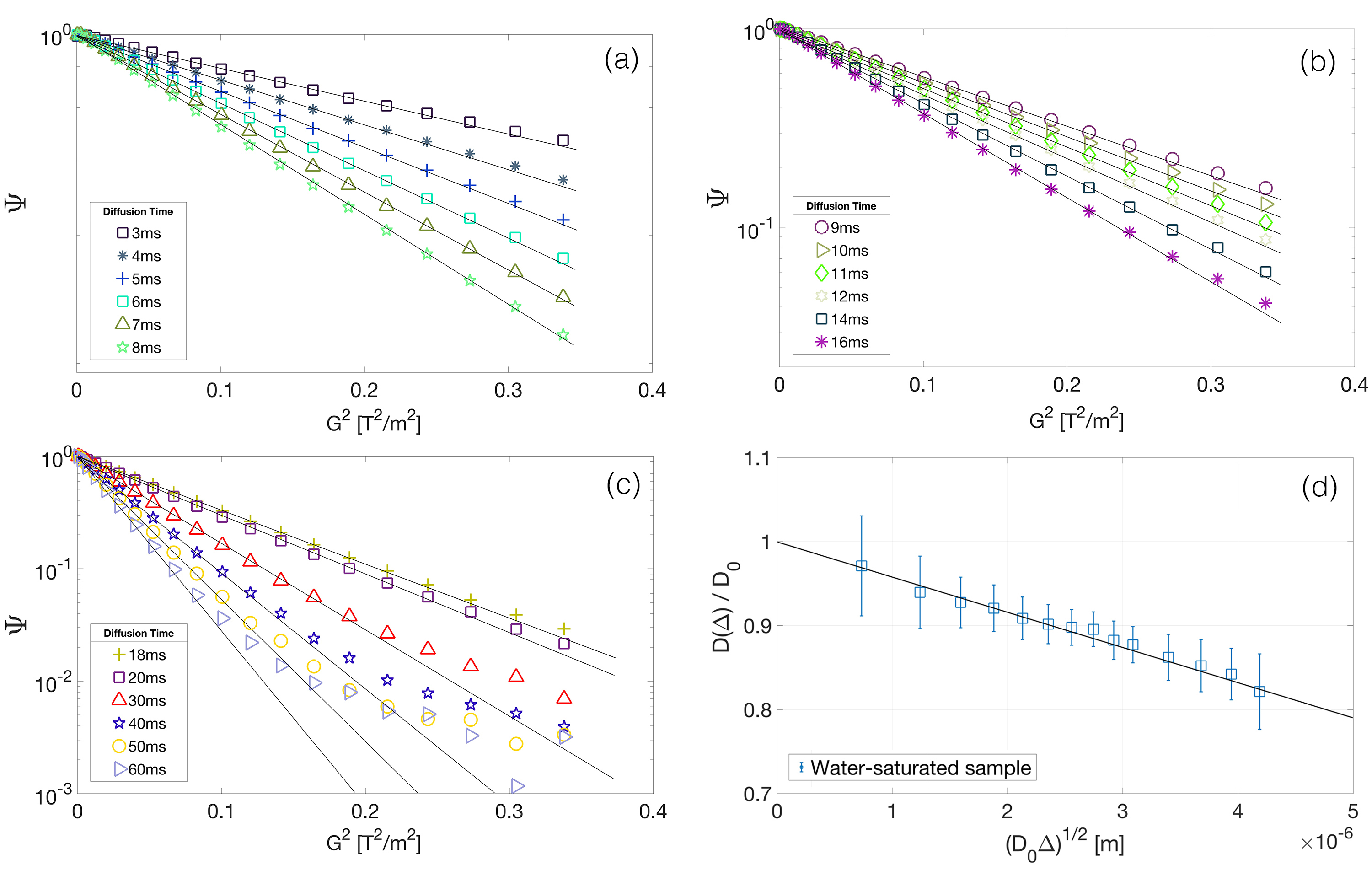}
		\caption{(a), (b) and (c) Signal attenuation $\Psi$ as a function of squared gradient $G^{2}$ for different diffusion times, on the water-saturated porous sample. Acquisitions were performed for 18 different $\Delta$ values between 3 and 60 ms. The black lines represent linear fittings considering the initial slope of each data set. Loss of linear behavior for different $\Delta$ values can be used to validate the Gaussian propagator approximation, and the short-time regime definition. (d) Analysis of the normalized time-dependent diffusion coefficients as a function of diffusion length $\sqrt{D_{0}\Delta}$, performed using the short-time approximation proposed by Mitra et al. In order to assure the validity of the short-time regime only the diffusion coefficients extracted from data sets with $\Delta$ values up to 20ms were considered. In plot (d) the black line represents a linear fitting considering only the term of order $\mathcal{O}[(D_{0}t)^{1/2}]$ in Equation (\ref{eq_Mitra}).}
		\label{fig_dif_water_rock_deff_fit}
	\end{center}
\end{figure*}

 Figure \ref{fig_dif_water_rock_deff_fit} shows signal attenuation $\Psi$ as a function of squared gradient $G^{2}$ for different diffusion times ((a), (b) and (c)) obtained from PFG measurements in the water-saturated porous sample, before oil injection. Observing signal attenuation in a vertical logarithmic scale it is possible to notice that the linear behavior, present in measurements with small $\Delta$ values ($<$ 20 ms), is gradually lost as diffusion time is increased. This graphical approach is a useful tool to evaluate the validity of the Gaussian propagator approximation, and to properly define a short-time regime for diffusion of a fluid inside a porous space. According to the behavior of signal decays showed in plots (a), (b) and (c) of Figure \ref{fig_dif_water_rock_deff_fit}, only the diffusion coefficients extracted from data sets with diffusion times up to 20 ms were considered for the analysis presented in the plot (d) of the same Figure, in order to ensure the validity of the Gaussian propagator approximation.\par  
 
 The plot (d) on Figure \ref{fig_dif_water_rock_deff_fit} shows an analysis of the extracted time-dependent diffusion coefficients as a function of the storage time $\Delta$, performed using the short-time approximation (Eq. (\ref{eq_Mitra})) proposed by Mitra et al. A linear fitting of the data was performed using the Curve Fitting Toolbox (MATLAB R2019b) considering initially only the first order term in Equation (\ref{eq_Mitra}). A sufficiently good adjustment was obtained with a surface-to-volume ratio value of $S/V =  9.22 \times 10^{4}$ m$^{-1}$ with an approximate error of 5 $\%$. The quality of the fit was verified by looking at the Sum of Squares due to Error (SSE) and the R-square coefficients, which are about 6$\times$10$^{-4}$ and 0.9896, respectively. The inclusion of the higher order term in the expansion for $D(t)$ adds two new parameters to the fit procedure, the surface relaxivity $\rho$ and the pore mean curvature $H$. In this new fit the values obtained for the new parameters $\rho$ and $H$ exhibited associated errors which were larger than the parameters' values themselves, whereas no improvement was observed in the fit quality coefficients and also no significant change was verified in the value of the  parameter $S/V$. Following the principle of simplicity, the first fit procedure, containing only the first order term in Equation (\ref{eq_Mitra}) , was considered. 
 
 An estimate of the surface-to-volume ratio for the porous space in a system composed by a random closed packing of mono-sized spheres can be performed using the relation $S/V = 6(1-\phi)/(d_{s}\phi)$, wherein $d_{s}$ stands for the spheres' diameter. Assuming $d_{s}$ = 250 $\mu$m and the close random packing porosity as $\phi \approx$ 37 $\%$ leads to a value of $S/V \approx  4.1 \times 10^{4}$ m$^{-1}$. Although the value obtained by NMR data fitting with Equation (\ref{eq_Mitra})) comes from an analytical expression, the estimated value on the other hand was obtained assuming an idealized system composed by a package of mono-sized spheres with perfect sphericity. The real measured system is the result of a sintering process performed in a random package of spheres with diameters varying within a size range, which are assumed to exhibit sphericities smaller than 1. A balance of all these factors is then expected to influence the resulting $S/V$ of the fabricated samples. For these reasons the estimated value was considered in fair agreement with the value extracted from PFG data using Equation (\ref{eq_Mitra}). 

The geometry of the samples used in this work can be fairly approximated by a packing of overlapping spheres, considering the effect of the applied sintering protocol in glass microspheres, analyzed in scanning electron microscopy (SEM) images \cite{Chencarek2019}. Weissberg \cite{WeissbergPermeability} deduced an approximate expression for the geometric tortuosity $\tau$ of overlapping spheres as a function of geometry porosity $\phi$:

\begin{equation}
\label{eq_tortuosity_spheres}
\tau_{g} = 1 - \frac{1}{2}ln\phi.
\end{equation}

 An estimate of samples permeability can be performed using the extracted $S/V$ value and the Kozeny-Carmen relation \cite{Schwartz1993Permeability}:

\begin{equation}
\label{eq_permeability}
k_{perm} \approx \Big(\frac{V}{S}\Big)^{2} \frac{\phi}{2\tau_{g}},
\end{equation}             

\noindent wherein $\phi$ is the porosity and $\tau_{g}$ represents porous media geometric tortuosity. A permeability estimate combining the $S/V$ value extracted from data in Figure \ref{fig_dif_water_rock_deff_fit}, and Equations (\ref{eq_tortuosity_spheres}) and (\ref{eq_permeability}), leads to a value of $k_{perm} \approx 16$ D with an approximate error of 12 $\%$, wherein in this case D stands for the permeability unit Darcy (1 D $ \approx$ 0.98 $\mu$m$^{2}$). The permeability value reported in \cite{Chencarek2019} for this sample, measured by the free gas expansion method, is $k_{perm} \approx 28$ D. It is important to remark that Equation (\ref{eq_permeability}) represents simply a correlation between permeability and porous media geometrical properties, and not an exact dependency. Therefore the permeability estimate performed with the surface-to-volume ratio value extracted from PFG data processing, and the tortuosity value obtained from Equation (\ref{eq_tortuosity_spheres}), can be considered in fair agreement with the measured value.    

\subsection{Post drainage PFG analysis}

\begin{figure*}[!ht]
	\begin{center}
		\includegraphics[width=\textwidth]{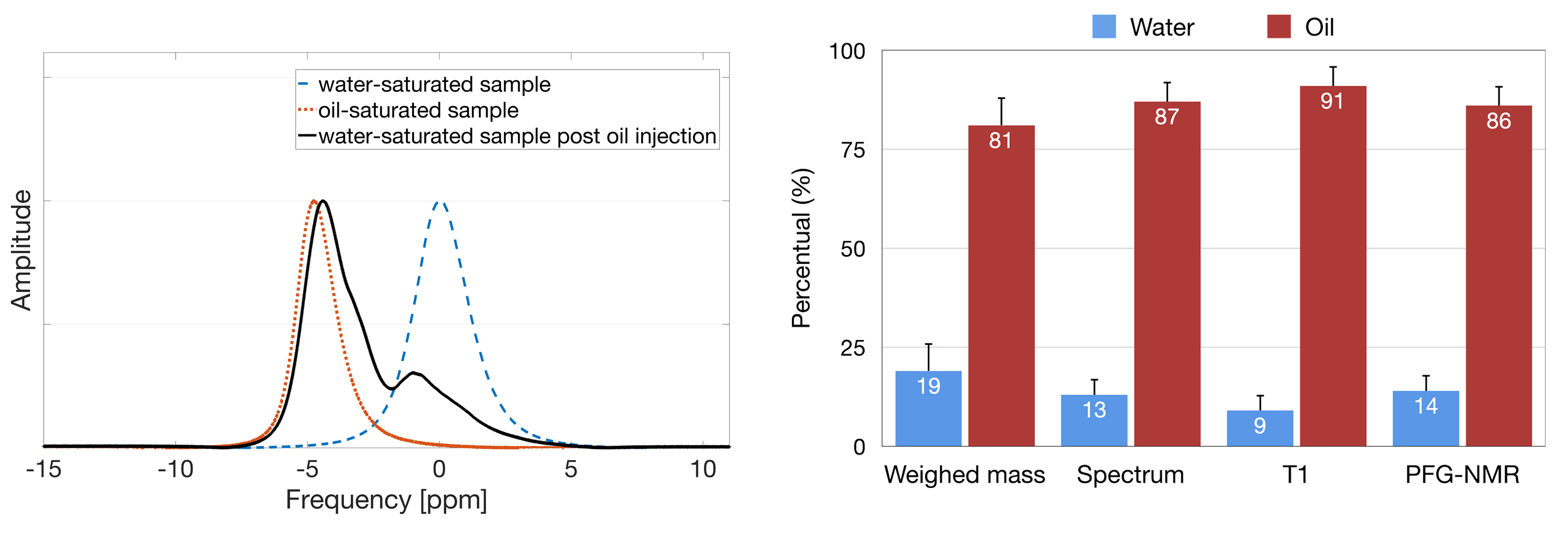}
		\caption{[Left] $^{1}$H spectra of water-saturated sample before and after drainage. [Right] Estimate of water and oil quantities obtained from weighed mass, spectroscopy, longitudinal relaxation ($T_{1}$) and PFG measurements after drainage. The estimate error (vertical black bars) was $6\%$ for the weighed mass method and $4\%$ for the NMR-based techniques.}
		\label{fig_post_injection_spectrum_amplitudes}
	\end{center}
\end{figure*}

$^{1}$H spectra of an oil-saturated sample, and a water-saturated sample before and after drainage, are shown in the left plot in Figure \ref{fig_post_injection_spectrum_amplitudes}. Although lines are particularly broad, it is possible to identify water and oil phases in the post drainage spectrum. A bi-Lorentzian fit was performed to estimate fluids proportion after drainage. In general, this evaluation should not be considered an accurate one, since distinct effects contribute to the observed spectrum line shapes. Considering the case of confined fluids, a line broadening in spectrum is expected due to the presence of grossly inhomogeneous magnetic fields inside porous matrix. In addition, NMR spectrum of crude oils present multiple peaks, relating to the combined chemical-shift structure of various molecules. In this case, the former mentioned line-broadening effect due to confinement will affect all peaks, and the shape of the resulting spectrum is not necessarily a Lorentzian. Here, the intent of the presented spectra is not to determine quantities of both fluids, but mainly to confirm the presence of both phases after drainage.\par 

Relative quantities of both phases were also evaluated by weighed mass, longitudinal relaxation ($T_{1}$) and PFG measurements (Fig. \ref{fig_post_injection_spectrum_amplitudes} - right plot). The model used to analyze the PFG data after drainage is discussed further. For the analysis of $T_{1}$ measurements after drainage both phases were assumed to be in the fast diffusion regime. Initial calibration experiments revealed that the fabricated samples saturated with a single fluid exhibit mono-exponential $T_{1}$ decays. Assuming that both phases after drainage are in fast diffusion regime and considering that there is no spin exchange between the water and the oil phases allows a simpler estimate for the relative content of both fluids using a bi-exponential model for the $T_{1}$ relaxation.    

For the case of water-saturated samples before drainage, Gaussian approximation for the diffusion propagator, and the extension of a short-time diffusion regime, can be evaluated through the analysis of signal attenuation linearity, in a semi-logarithmic $\Psi(t)$ versus squared gradient $G^{2}$ plot. However, after oil injection, in the presence of two fluids with different diffusion coefficients, PFG signal attenuation will no longer exhibit a single Gaussian behavior. Although a graphical validation for this approximation is no longer possible, the hypothesis that measurements may be carried in a short-time regime, where the Gaussian propagator approximation could be considered valid for both fluids, is quite reasonable, and is assumed in the presented analysis. \par

A two-fluid model was proposed by Stallmach and Thomann \cite{PatentStallmachDiffusion} to determine fluid fractions with different translational mobilities in porous media, by PFG NMR measurements. The analysis of time-dependent diffusion coefficients from water-saturated samples after drainage was carried considering a similar bi-Gaussian model in the form:

\begin{equation}
\label{eq_bi_Gaussian}
\begin{aligned}
\Psi(t_{e}) = {} & s_{w}exp[-D^{w}\gamma^2(A_{p}(t_{e}))] \\
                      & + s_{o}exp[-D^{o}\gamma^2(A_{p}(t_{e}))],
\end{aligned}
\end{equation}

\noindent in which $D^{w}$, $D^{o}$, $s_{w}$ and $s_{o}$ denotes water and oil diffusion coefficients and their respective saturations, with $s_{w} + s_{o} = 1$, and $A_{p}(t_{e})$ has the form expressed in Equation (\ref{dif_attenuation_sol_pulsed}). \par 

\begin{figure*}[!ht]
	\begin{center}
		\includegraphics[width=\textwidth]{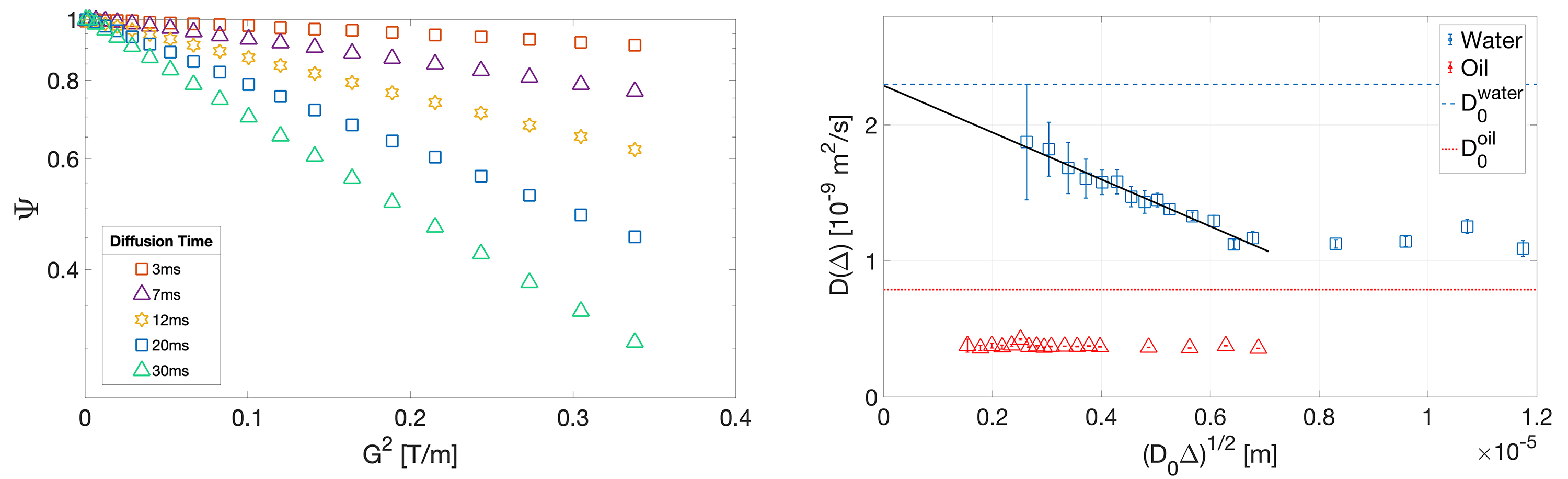}
		\caption{[Left] Signal attenuation $\Psi$ as a function of squared gradient $G^{2}$ for different diffusion times, on the water-saturated porous sample after drainage. Acquisitions were performed for $18$ different $\Delta$ values between 3 and 60 ms. Representative signal decays for $\Delta$ values 3, 7, 12, 20 and 30 ms are shown. [Right] Time-dependent diffusion coefficients obtained for water (blue squares) and oil (red triangles) extracted using the bi-Gaussian model in Equation (\ref{eq_bi_Gaussian}) from PFG experiments in water-saturated sample after drainage. Fitting error bars are shown for both fluids, however for the oil points error is smaller than marker size. Dashed blue line and dotted red line represent bulk values of water and oil diffusion coefficients ($D^{w}_{0}$ and $D^{o}_{0}$), respectively. The black line on the right plot represents the analysis of post drainage water time-dependent diffusion coefficients as a function of diffusion length $\sqrt{D_{0}\Delta}$, performed using the short-time approximation (Eq. (\ref{eq_Mitra})) considering only the term of order $\mathcal{O}[(D_{0}t)^{1/2}]$. A new water-probed surface-to-volume ratio $S/V =  3.01 \times 10^{5}$ m$^{-1}$ was found, representing an increase of more than 300 $\%$ compared to its value before drainage.}
		\label{fig_deff_Delta_injection_water_analysis}
	\end{center}
\end{figure*}

Representative signal decays obtained from post drainage PFG measurements for five different $\Delta$ values are shown in the left plot of Figure \ref{fig_deff_Delta_injection_water_analysis}. The right plot in Figure \ref{fig_deff_Delta_injection_water_analysis} shows the time-dependent diffusion coefficients obtained from all the 18 performed measurements for water and oil, as a function of the diffusion length ($\sqrt{D_{0}\Delta}$), from PFG data analysis after drainage using Equation (\ref{eq_bi_Gaussian}). Throughout the $\Delta$ time interval investigated water diffusion coefficient values (blue squares) showed a clear attenuation while diffusion time is increased. As expected in both scenarios illustrated in Figure \ref{fig_injection}, wetting phase exhibits a good connectivity, as water molecules probe the new available porous space after drainage. A new analysis of the short-time behavior of water time-dependent diffusion coefficients (black line in the right plot) using Equation (\ref{eq_Mitra}) reveals a new water-probed surface-to-volume ratio $S/V =  3.01 \times 10^{5}$ m$^{-1}$; an increase of more than 300 $\%$ compared to its value before oil injection. This behavior is expected since not only the surface area of the water phase is increased with the addition of a contact surface between wetting and non-wetting phases, but also a significant amount of wetting phase is drained in the injection to be occupied by the non-wetting phase (Fig.  \ref{fig_post_injection_spectrum_amplitudes}).         

Oil time-dependent diffusion coefficients (red triangles - right plot of Fig. \ref{fig_deff_Delta_injection_water_analysis}) nevertheless showed no significant variation with diffusion time. Such behavior can be analyzed over two different perspectives. In the first one, oil phase would also be distributed in a connected configuration, and considering that oil molecules bulk diffusivity is approximately three times smaller than water (Fig. \ref{fig_dif_water_oil_bulk}), the diffusion time range investigated, from 3 to 60 ms, would not be long enough for one to measure an expressive attenuation in oil $D(\Delta)$ values, similar to the one observed in water during the same time range. Although a low diffusivity could be used to justify the small variation between oil $D(\Delta)$ values, the time-dependent diffusion coefficient of oil observed in a post drainage PFG experiment with the shortest available diffusion time ($\Delta = $ 3 ms) corresponds to almost half of its bulk value (red dotted line - left plot of Fig. \ref{fig_deff_Delta_injection_water_analysis}). This considerable attenuation, observed during a small diffusion time, would not be expected assuming that oil is distributed in a connected configuration.\par

A different analysis can be done considering that oil diffusion coefficients have actually reached a stationary regime, which, in this case, indicates a highly-restricted diffusion process. Here, a scenario where non-wetting phase is poorly connected would be most likely to describe fluids conformation after drainage, and a distribution of oil-in-water droplets may have formed within pores. An expression for the signal attenuation of molecules diffusing inside spherical cavities in PFG experiments was calculated by Murday and Cotts \cite{MurdayCottsDroplets1968}, and an approximation for it was later proposed by Callaghan et al. \cite{CALLAGHAN1983Cheese}. Assuming that oil droplets have spherical shape, and that a Gaussian distribution of droplets radii is to be found along the sample, PFG signal attenuation due to oil molecules diffusion can be approximated\footnote{The validity of this approximation is based on the assumption that, although short, gradient pulses have a finite duration $\delta$, and spins, during this interval, accumulate a Gaussian distribution of phases. The correct corresponding between $r_{0}$, $\sigma$ and signal attenuation, using Equation (\ref{eq_droplet_dist_Callaghan}), should be obtained for small values of $\beta^{2}$ \cite{CALLAGHAN1983Cheese}. } by \cite{CALLAGHAN1983Cheese,McDonald1999Droplet}:

 \begin{equation}
 	\Psi(\delta, G, r_{0}, \sigma) = \Psi_{0} \frac{1}{\sqrt{1+2\sigma^{2}\beta^{2}}}exp\Bigg(-\frac{\beta^{2}r^{2}_{0}}{1+2\sigma^{2}\beta^{2}}\Bigg),
 	\label{eq_droplet_dist_Callaghan}
 \end{equation}        

\noindent wherein $r_{0}$ is the average radius of the droplets radii distribution, $\sigma$ is the standard deviation and $\beta^{2} = \gamma^{2}\delta^{2}G^{2}/5$.  

\begin{figure}[!ht]
	\begin{center}
		\includegraphics[width=220pt]{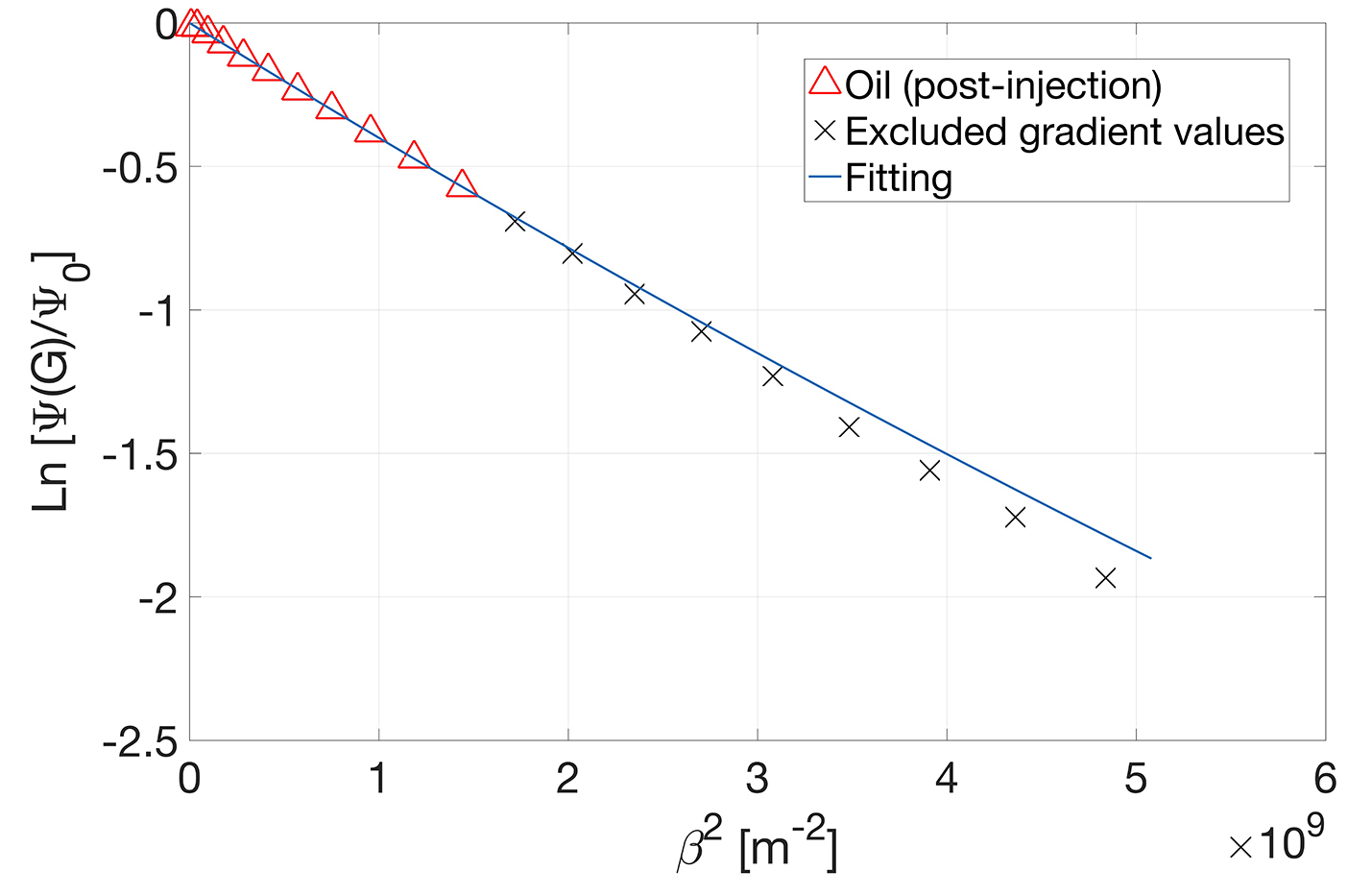}
		\caption{Signal attenuation $ln[\Psi(\delta, G, r_{0}, \sigma) / \Psi_{0}]$ as a function of $\beta^{2}$ for oil diffusion after drainage. Blue solid line denotes the fitting performed with Equation (\ref{eq_droplet_dist_Callaghan}) considering the formation of oil-in-water droplets within pores. Validity of the carried approximations can be observed through the agreement between data and the proposed model (red triangles) for small $\beta^{2}$ values. Cross markers represent large $\beta^{2}$ values that were excluded from fit. The extracted values for the average radius and the standard deviation assuming a Gaussian distribution of oil droplets radii were $r_{0} \approx$ 20 $\mu$m and $\sigma \approx$ 4 $\mu$m, respectively. The error in the estimate of $r_{0}$ was less than 1 $\%$. } 
		\label{fig_fit_droplet_Callaghan_Gauss_dist}
	\end{center}
\end{figure}

Once the diffusion time $\Delta$ was increased to the regime where signal attenuation $ln[\Psi(\delta, G, r_{0}, \sigma) / \Psi_{0}]$ does not depend on $\Delta$, the plot of $ln[\Psi(\delta, G, r_{0}, \sigma) / \Psi_{0}]$ versus $\beta^2$ can be adjusted with Equation (\ref{eq_droplet_dist_Callaghan}) for small values of $\beta^2$, and assuming a Gaussian distribution of droplets radii, $r_{0}$ and $\sigma$ values can be estimated directly from data \cite{CALLAGHAN1983Cheese}. Figure \ref{fig_fit_droplet_Callaghan_Gauss_dist} presents the signal attenuation $ln[\Psi(\delta, G, r_{0}, \sigma) / \Psi_{0}]$ as a function of $\beta^{2}$, and a fitting performed with Equation (\ref{eq_droplet_dist_Callaghan}) for oil diffusion after drainage. Validity of the carried approximations can be observed through the agreement between data and the assumed model for small $\beta^{2}$ values. Extracted values for the average radius and the standard deviation were $r_{0} \approx$ 20 $\mu$m and $\sigma \approx$ 4 $\mu$m, respectively. 

Although the bi-exponential model employed in the analysis of PFG data is based in a set of a priori information on the investigated system, which from the physical standpoint upholds the interpretation of the results, its application relies in a mathematical fit procedure. Grebenkov \cite{Grebenkov2017Chapter3DiffusionBook} addressed the mathematical limitations of the bi-exponential model on the analysis and interpretation of effective diffusion coefficients obtained from diffusion measurements in a two-phase system. On what regards the fit procedure, the analysis of PFG decays obtained for small diffusion times are the most challenging, as for these curves the signal attenuation is considerably smaller than the one obtained for larger diffusion times, as shown in the left plot of Figure \ref{fig_deff_Delta_injection_water_analysis}. It must be considered the possibility that oil diffusion coefficients obtained from bi-exponential fits performed in the data sets presented in this work may be misrepresented for the case of small diffusion times. The option of analyzing such data sets by inverse Laplace transform was discarded considering that the procedure is also affected by nonzero baseline offsets \cite{istratov1999exponential}. Nevertheless, water diffusion coefficients obtained from the same fit procedure exhibited a very consistent behavior with respect to what was expected physically for the conformation of both fluids after drainage.

\section{Conclusion}

Time-dependent diffusion coefficients measured by PFG NMR can be used to characterize drainage experiments, providing information on the individual phases and post drainage fluid conformation. Pre drainage PFG measurements in water-saturated samples were used to extract confinement features, and estimates of samples surface-to-volume ratio and permeability values, carried from PFG data analysis, were shown to be in fair agreement with analytical and reported results, respectively. The short-time analyses of time-dependent diffusion coefficients obtained from PFG measurements can be used to characterize the increase in surface-to-volume ratio probed by the wetting phase after drainage.\par   

Wetting and non-wetting phase time-dependent diffusion coefficients, extracted from PFG NMR experiments, can be analyzed to infer dynamics of single phases and to portray post drainage fluids conformation scenarios. The case where non-wetting phase was considered to exhibit a poorly connected geometry was analyzed assuming a restricted diffusion process and the formation of a oi-in-water distribution of droplets within pores, and PFG signal attenuation was used to determine a Gaussian distribution of oil-in-water droplets radii. \par 

Analyses of post drainage PFG measurements were performed using simple bi-Gaussian models. Although the presented analysis required a particular set of approximations regarding self-diffusion regimes and diffusion propagators that are, nevertheless, common in the analysis of PFG NMR measurements in confined systems, data behavior itself can be used to check the regime in which said approximations become valid, so to ensure a proper interpretation of raw data and better estimations of determined parameters. \par

\section*{Declaration of competing interest}

The authors declare that they have no known competing financial interests or personal relationships that could have appeared to influence the work reported in this paper.

\section*{CRediT authorship contribution statement}

\textbf{Bruno Chencarek:} Conceptualization, Methodology, Investigation, Writing - Original Draft, Visualization. \textbf{Moacyr Nascimento:} Conceptualization, Validation, Writing - Review $\&$ Editing. \textbf{Alexandre Martins Souza:} Writing - Review $\&$ Editing. \textbf{Roberto S. Sarthour:} Writing - Review $\&$ Editing. \textbf{Bernardo Coutinho C. dos Santos:} Conceptualization, Validation, Writing - Review $\&$ Editing. \textbf{Maury Duarte Correia:} Conceptualization, Validation, Writing - Review $\&$ Editing, Funding acquisition. \textbf{Ivan dos Santos Oliveira:} Supervision, Project administration, Writing - Review $\&$ Editing. 

\section*{Acknowledgment}

The authors would like to thank Brazilian research agencies CAPES and CNPq for the support. We also greatly appreciate all the enlightening comments and points raised by the anonymous referees during the revision of the manuscript. Authors also acknowledge Marcel Moura, from the Physics Department of University of Oslo, for discussions on fluid dynamics and drainage experiments. This work was funded by PETROBRAS through a Research and Development Project with the Brazilian Center for Research in Physiscs (project \textnumero \, 2017/00486-1).

\printbibliography

\end{document}